\documentclass[aps,prl,reprint,superscriptaddress,nofootinbib,longbibliography]{revtex4-2}
\usepackage{amsmath,amssymb,bm,mathtools}
\usepackage{graphicx}
\usepackage{microtype}
\usepackage[colorlinks=true,citecolor=blue,linkcolor=blue,urlcolor=blue]{hyperref}
\usepackage{booktabs}
\newcommand{\dd}{\mathrm d}
\newcommand{\e}{\mathrm e}
\newcommand{\diag}{\operatorname{diag}}
\newcommand{\Var}{\operatorname{Var}}
\newcommand{\T}{\mathsf T}
\newcommand{\chisq}{\chi^2}
\newcommand{\KL}{D_{\mathrm{KL}}}
\newcommand{\ymin}{y^{\min}_e}

\newcommand{\p}{\bm p}
\newcommand{\pii}{\bm\pi}

\newcommand{\ket}[1]{\lvert #1\rangle}
\newcommand{\bra}[1]{\langle #1\rvert}
\newcommand{\braket}[2]{\langle #1\vert #2\rangle}
\newcommand{\mel}[3]{\langle #1\vert #2\vert #3\rangle}

\begin{document}

\title{Slow is fast: raising barriers to accelerate thermal relaxation}

\author{Yubo Wang}
\affiliation{Department of Chemistry, University of North Carolina at Chapel Hill, Chapel Hill, North Carolina 27599, USA}
\author{Zhiyue Lu}
\email{zhiyuelu@unc.edu}
\affiliation{Department of Chemistry, University of North Carolina at Chapel Hill, Chapel Hill, North Carolina 27599, USA}

\date{\today}

\begin{abstract}
For a reversible system relaxing to equilibrium, the obvious fastest strategy is to lower all kinetic barriers (open all gates). We find that such intuition holds at three levels: the all-open-gate strategy achieves the highest local conductance, it maximizes the instantaneous speed of approach in every $f$-divergence, and it simultaneously maximizes all relaxation eigenvalues. Nevertheless, we show that a counter-intuitive finite-time optimum lies beyond this intuition and operates at a fourth level, invisible to all three: eigenvector rotation. Noncommutativity enables timed schedules to reproject residual amplitudes across relaxation modes, thereby achieving faster relaxation. Optimal schedules are bang--bang. In our illustrative example, the best-found schedule also employs counter-gating, transiently raising selected barriers, and reduces the terminal residual by a factor of $130$ relative to all-open, and by $7$ relative to the best static landscape. A no-go theorem shows that noncommutativity is necessary: commuting generators collapse every schedule to a static time-averaged landscape, worse than the intuitive static control. In the reverse problem, the dual schedule preserves nonequilibrium free energy far more effectively than intuitively keeping all barriers at maximum heights. Whether accelerating or delaying relaxation, barrier control performs no work on the reduced Markov system; it only re-times a fixed total dissipation budget.
\end{abstract}

\maketitle

Which of two states reaches equilibrium first, the closer one or the farther one?  A decade of Markovian Mpemba-effect studies has shown that \emph{the intuitive answer can fail}, because far-from-equilibrium relaxation is governed not by ``distance'' alone but by how the initial state decomposes onto the relaxation modes of the generator: an initial condition that avoids the slow modes outruns one that does not \cite{LuRaz2017,Klich2019,Kumar2020,Bechhoefer2021}. In all such effects, however, the Mpemba advantages are \emph{inherited}: fixed simply by the initial state or by a fixed generator that, once chosen, never changes during relaxation. Here we ask a different question: during relaxation itself, and without doing work or altering the equilibrium, can the barriers be scheduled to speed it up or slow it down? The answer reveals an apparent paradox. The intuitive prescription, i.e., set every barrier to its extreme, provably gives the optimal transient relaxation speed at all times, and it maximizes every relaxation eigenvalue (optimal speed for each eigenmode). However, a counter-intuitive schedule that at times \emph{raises} some barriers to maximally high can reach equilibrium sooner. The resource that beats the intuitive method is not a cleverer initial state or instantaneous rate (as in the Mpemba effect), but the \emph{time ordering} of noncommuting relaxation generators that all share the same equilibrium. 

In this Letter, barrier heights are a special control: they set how fast a system relaxes without changing its final equilibrium, and, within the reduced Markov description, tuning them changes no state energy and thus performs no system work. Controlling barriers is also achievable in realistic processes: in catalysis, gate-controlled mesoscopic systems, and stochastic pump models, kinetic prefactors or transition-state energies can be tuned while the state energies, and hence the equilibrium distribution, remain fixed \cite{Kramers1940,Hanggi1990,SinitsynNemenman2007,Rahav2008,Chernyak2008}. Barrier-only control therefore isolates the question sharply: with both the initial and final states (distributions) fixed, a control schedule can act only on the \emph{route}. This is precisely the finite-time-control question that Mpemba-effect studies leave open.

The mechanism reported in this Letter is distinct from familiar routes to rapid relaxation. In the literature, engineered swift equilibration reshapes the energy landscape, while thermodynamic-geometry, minimum-work, and optimal-transport approaches optimize the cost of a prescribed driven transformation \cite{Martinez2016,GueryOdelin2019,Plata2021,Schmiedl2007,Aurell2011,Sivak2012,Zulkowski2012,Prados2021}. Thermal quenches, state-preparation protocols, and Mpemba-like effects supply the favorable modal amplitudes \emph{by hand}, through the initial condition or a fixed generator \cite{Pemartin2024,BeatoTeza2026,LuRaz2017,Klich2019,Kumar2020,Kumar2022,Bechhoefer2021}. Other approaches such as the nonreversible Markov-chain accelerations gain speed by breaking detailed balance or leaving the reversible class \cite{Ichiki2013,Kaiser2017}.

Three recent directions sit closest to this Letter, and the contrast with them precisely motivates the present mechanism. First, relaxation can be accelerated by a genuinely \emph{time-dependent} protocol: precooling a system before heating it yields exponentially faster heating \cite{GalRaz2020,Chittari2023}. There, however, the schedule acts through the bath temperature or chemical concentration, so the instantaneous equilibrium itself moves and heat or chemical work is exchanged along the way; the gain in \cite{GalRaz2020} is an amplitude prepared by a quench, not a consequence of time ordering. Second, relaxation timescales can be set by applying a unitary operation to the initial state that cancels chosen relaxation modes, with the generator held fixed throughout \cite{BeatoTeza2026}; in this case, the modal amplitudes are fixed once, before the dynamics begins. Third, the structure of the system-bath coupling itself reshapes the relaxation spectrum \cite{TezaBoundary2023}, but the resulting generator is again held fixed in time. In this Letter, by contrast, the initial state and the final equilibrium are both fixed, every accessible generator obeys detailed balance with respect to that same equilibrium, and the only freedom is the time-dependent control over the barrier height of reversible kinetic channels. The advantage is therefore created \emph{during} relaxation, by the order in which reversible channels are opened (minimum barrier) and closed (maximum barrier). 

\emph{Barrier-controlled detailed balance.---}
Consider a connected graph whose vertices are states with energies $E_i$ and whose edges are transitions. The corresponding equilibrium distribution follows the Boltzmann form $\pi_i\propto\e^{-\beta E_i}>0$, with a fixed inverse temperature $\beta=(k_BT)^{-1}$. An undirected edge $e=\{i,j\}$ carries a symmetric conductance set by its barrier height $B_e$:
\begin{equation}
 g_e(t)=\exp[-\beta B_e]=\bar g_e y_e(t),\qquad y_e(t)\in[\ymin,1],                 \label{eq:conductance}
\end{equation}
with transition rates $k_{i\leftarrow j}=g_e/\pi_j$ and $k_{j\leftarrow i}=g_e/\pi_i$. Regardless of the choice of $B_e$ (or effectively $y_e$), the corresponding generator obeys detailed balance with respect to the \emph{same} $\pii$. The control of barrier heights is parameterized by
\[
y_e=\exp[-\beta(B_e-B_e^{\min})] >0,
\]
which linearly modifies the conductance $g_e$. The highest conductance achievable at open-gate is denoted by $\bar g_e = \exp(-\beta B_e^{\min})$. We call $y_e=1$ the ``open gate'' setting, as it puts the barrier at its minimum height, and $y_e=\ymin$ the ``closed gate'' setting, as it puts the barrier at its maximum height. The relative entropy between any distribution and the final Boltzmann distribution satisfies $\KL[\p\Vert\pii]=\beta(F[\p]-F[\pii])$, so it measures the nonequilibrium free energy stored in the deviation from equilibrium \cite{Qian2001,VaikuntanathanJarzynski2009,EspositoVdB2011}.

Our only handle is the barrier schedule $\{y_e(t)\}$: the target distribution cannot be moved and no work can be done, so the sole freedom is the kinetic timing.

To make ``fast'' and ``slow'' precise, we need to measure distance from equilibrium. Our primary analytical distance is the chi-square divergence,
$
\chisq[\p\Vert\pii]=\sum_i\frac{(p_i-\pi_i)^2}{\pi_i},
$
because it turns the relaxation problem into a symmetric linear spectral one. We also consider the relative entropy,
$
\KL[\p\Vert\pii]=\sum_i p_i\ln(p_i/\pi_i),
$
which equals the nonequilibrium free-energy difference in units of $k_{\rm B}T$ and provides the thermodynamic interpretation. More generally, we consider convex $f$-divergences \cite{AliSilvey1966,Csiszar1967},
$
D_f=\sum_i\pi_i f(p_i/\pi_i),
$
where $f$ is convex and satisfies $f(1)=0$, a family recently used to sharpen notions of relaxation speed in the Mpemba effect \cite{VanVuHayakawa2025,Teza2025}. For a smooth $f$ and $\delta p=p-\pi$ sufficiently small,
$
D_f=\tfrac12 f''(1)\chisq+O(\|\delta p\|^3).
$
so $\chisq$ is the universal quadratic form underlying all smooth $f$-divergences. We therefore develop the theory using $\chisq$, while using $\KL$ to connect the results to stochastic thermodynamics.
Symmetrizing the residual by the standard similarity transformation \cite{vanKampen},
\begin{equation}
 \ket q=\Pi^{-1/2}(\ket p-\ket\pi),\qquad
 \braket q q=\chisq[\p\Vert\pii],                                \label{eq:qdef}
\end{equation}
with $\Pi=\diag(\pii)$, turns the master equation into
\begin{equation}
 \ket{\dot q}=-L[y]\ket q,\qquad
 L[y]=\sum_e y_eL_e,\qquad L_e=\bar g_e\ket{v_e}\!\bra{v_e},      \label{eq:dynamics}
\end{equation}
a sum of one positive-semidefinite rank-one operator per edge, with $\ket{v_e}=\Pi^{-1/2}(\ket i-\ket j)$. Each $L_e\succeq0$ and annihilates the common null vector $\ket{\sqrt\pi}$, the edge/network structure being that of Schnakenberg \cite{Schnakenberg1976}. Every question below is thus a question about the time-ordered product of the positive-semidefinite operators $L[y(t)]$. The \emph{structural} results below, i.e., monotone decay, bang--bang optimality, and the no-go theorem, all hold for every convex $f$-divergence. 

\emph{Why the intuitive method appears optimal.---}
The intuitive control holds every gate at its extreme: $y_e\equiv1$ (all barriers set to lowest) to hurry, $y_e\equiv\ymin$ (all
barriers set to highest) to stall. This intuition is independent of the energy landscape or initial state, and it is fixed (time-independent). This claim to optimal relaxation speed is not naive: It can be defended at three successively stronger levels. \emph{(1)} It is the locally optimal choice for each edge: barriers impede motion, so fully lowering a barrier maximally speeds relaxation through that edge, and fully raising it maximally delays relaxation through that edge. \emph{(2)} It is the transiently optimal choice for \emph{every} convex $f$-divergence at all times. With $x_i=p_i/\pi_i$,
\begin{equation}
 -\frac12\frac{\dd}{\dd t}\chisq
 =\mel q{L[y]}q
 =\sum_{e=\{i,j\}}\bar g_e y_e(x_i-x_j)^2\ \ge0,                
\label{eq:dirichlet}
\end{equation}
and for a general convex $f$-divergence $-\dot D_f=\sum_{e=\{i,j\}}\bar g_e\,y_e\,(x_i-x_j)[f'(x_i)-f'(x_j)]\ge0$ because $f'$ is nondecreasing (one-line derivations in \cite{Supplement}). Every convex $f$-divergence thus decreases monotonically along every schedule, and its decay rate is, term by term, \emph{linear and monotonic in each control} $y_e$ with a nonnegative coefficient: at every instant and in every state,
the extremes maximize (or minimize) the instantaneous rate of approach.   \emph{(3)} The intuition is confirmed at the level of the whole spectrum. Writing $L[y]\ket{u_a}=\gamma_a\ket{u_a}$, Weyl monotonicity \cite{Bhatia1997} implies that opening barriers maximizes  \emph{every} ordered relaxation rate, since $L_{\rm open}-L[y]=\sum_e(1-y_e)\bar g_e\ket{v_e}\bra{v_e}\succeq0$. Steepest at every instant, uniformly over states, and mode by mode across the spectrum: the extremes appear unbeatable. We keep them as the benchmark. It is worth to point out the distinction from existing studies in optimal control: That a greedy policy can be beaten is generic in optimal control. What is not generic is that all-open carries \emph{three independent optimality certificates}---pointwise in every convex $f$-divergence, in every state, and across the entire spectrum---any one of which would appear to settle the time-independent problem, and that it nonetheless fails at finite time.

\emph{Yet transiently justified intuition fails.---}
We now state the goal precisely: to minimize (or maximize) the distance to equilibrium at a finite final time $T$. All three arguments above then miss the decisive variable. For a static landscape, the final distance to equilibrium reads
\begin{equation}
 \chisq(T)=\sum_{a\ge1}|\braket{u_a}{q_0}|^2\e^{-2\gamma_aT}.    \label{eq:staticmodes}
\end{equation}
So the terminal distance is set \emph{jointly} by the relaxation eigen-rates $\gamma_a$ and the initial modal decomposition amplitudes $\braket{u_a}{q_0}$. The intuitive method optimizes every eigen-rate but neglects the decomposition amplitudes, and the two are coupled through the eigenvectors. Because the edge generators $L_e$ generically do not commute, changing a gate $y_e$ rotates the eigenvectors at the same time as it shifts the eigenvalues. Merely optimizing all the eigen rates can therefore adversely load amplitude onto the slow modes: the highway is widened, yet the traffic is routed onto the slow lanes. This is the Mpemba lesson reappearing as the cost of transiently greedy control: there, a farther initial state wins by avoiding the slow modes of a fixed generator \cite{LuRaz2017,Klich2019,TezaCrossing2023}; here, a landscape with uniformly slower modes wins by placing the residual better. The optimal static strategy still belongs to the eigen-analysis tradition of the Mpemba effect; the time-ordered control extends that logic from state selection to kinetic scheduling.

The noncommutativity is not incidental but \emph{necessary} for counter-intuitive controls of relaxation. If all $L_e$ commute, time ordering disappears,
\begin{align}
 \mathcal T\exp\!\left[-\int_0^TL[y(t)]\dd t\right]
 &=\exp[-TL[\bar y]],\nonumber\\
 \bar y_e&=T^{-1}\!\int_0^Ty_e(t)\dd t,                           \label{eq:commuting}
\end{align}
simultaneous diagonalization makes every modal factor monotone in every $\bar y_e$, and the intuitive fixed-extreme strategy is globally optimal against arbitrary time dependence. Only when the accessible generators fail to commute can any schedule beat it (End Matter A).

\emph{Steering the amplitudes.---}
Mpemba-type effects inherit their amplitude structure from the initial
condition. A time-dependent schedule can instead \emph{reproject} the
evolving residual among the decaying modes---evacuating the slow modes
to accelerate, or loading them to delay. A static landscape selects
its eigenframe only once, whereas a schedule can repeatedly reproject
among different modal bases. With $\ket n=\ket q/\|q\|$ and the effective contraction rate $\gamma_{\rm eff}=\mel nLn=-\tfrac12\dot\chisq/\chisq$, a static landscape only \emph{filters},
\begin{equation}
 \dot\gamma_{\rm eff}=-2\Var_n(L)\le0,                             \label{eq:filtering}
\end{equation}
monotonically depleting the fast modes until a slow mode dominates. Allowing the eigenframe to move adds rate drift and eigenframe reprojection: in the instantaneous eigenbasis $L=U\Gamma U^{\T}$, with $\hat c$ being the normalized residual and connection being $A=U^{\T}\dot U=-A^{\T}$,
\begin{equation}
 \dot\gamma_{\rm eff}
 =\underbrace{\mel{\hat c}{\dot\Gamma}{\hat c}}_{\text{rate drift}}
  \underbrace{-2\Var_{\hat c}(\Gamma)}_{\text{filtering}\,\le0}
  \underbrace{-\mel{\hat c}{[\Gamma,A]}{\hat c}}_{\text{reprojection}},   \label{eq:threeterm}
\end{equation}
derived in End Matter B. 
At fixed instantaneous spectrum, the reprojection term is the only
contribution that can oppose static filtering by redistributing modal
amplitudes. It can be nonzero only when the eigenframe moves; such
motion is unavailable when all controlled generators commute. The amplitude structure that Mpemba-type effects must initialize with is therefore synthesized here on demand, during the relaxation itself. Unlike nonreversible accelerations, which gain speed by breaking detailed balance and thus leave the reversible class \cite{Ichiki2013,Kaiser2017}, this reprojection is achieved entirely within it: every instantaneous generator remains reversible with respect to the same equilibrium.

\begin{figure*}[t]
\includegraphics[width=\textwidth]{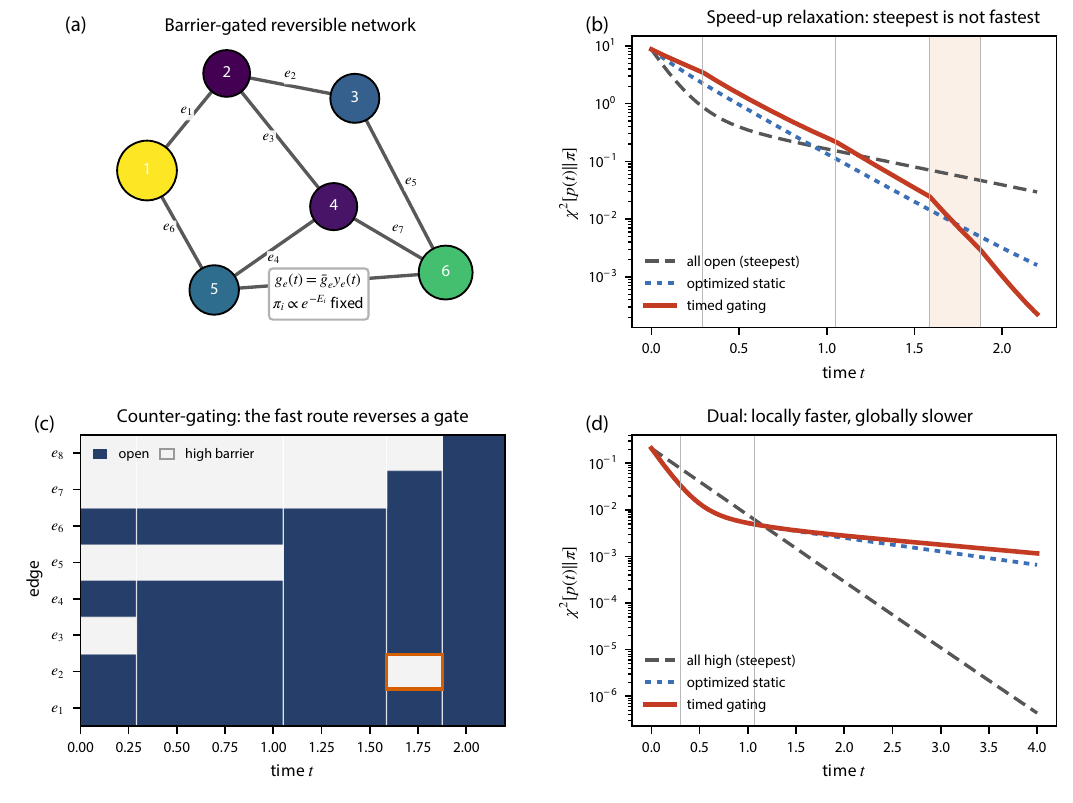}
\caption{\textbf{Barrier scheduling outperforms the pointwise
extremes.}
(a) Six-state reversible network; barrier control changes the
conductances while leaving $\pii$ fixed.
(b) Acceleration at $\ymin=0.05$ and $T=2.2$.
Although all-open maximizes the instantaneous decay rate at every fixed
state, timed gating reduces the terminal $\chisq$ by factors of $130$
and $7.05$ relative to all-open and the best-found static landscape,
respectively. The shaded interval marks the nonmonotone reversal shown
in panel (c).
(c) Acceleration gate schedule. Edge $e_2$ is opened, driven to the
high-barrier extreme, and reopened, producing an explicit counter-gate
rather than a monotone approach to all-open. The orange box highlights
the temporary high-barrier interval of $e_2$, the middle step of the
open--high--open reversal.
(d) Delay at $\ymin=0.30$ and $T=4$. Opening selected barriers before
an all-high hold retains $2.67\times10^3$ and $1.76$ times as much
terminal residual as the all-high protocol and the best-found static
landscape, respectively.}
\label{fig:main}
\end{figure*}

\emph{The optimum is bang--bang.---}
Pontryagin's minimum principle \cite{Pontryagin1962,BechhoeferControl2021} supplies necessary switching conditions and guides the numerical search. For terminal cost $\Phi=\braket{q(T)}{q(T)}$, introduce the adjoint
\begin{align}
 \ket{\dot\eta}&=L[y]\ket\eta,\qquad \ket{\eta(T)}=2\ket{q(T)},\nonumber\\
 G_e(t)&=\mel\eta{L_e}q,                                           
\label{eq:adjoint}
\end{align}
whose score $G_e$ weighs each barrier's instantaneous value \emph{to the terminal objective}; a first variation gives
\begin{equation}
 \delta\Phi=-\int_0^T\sum_eG_e(t)\,\delta y_e(t)\dd t.           \label{eq:variation}
\end{equation}
Because the control enters linearly, nonsingular optima are bang--bang \cite{Pontryagin1962,Liberzon2012,Schaettler2012}:
\begin{equation}
\begin{array}{c|cc}
 &G_e>0&G_e<0\\ \hline
\text{minimize }\Phi&y_e=1&y_e=\ymin\\
\text{maximize }\Phi&y_e=\ymin&y_e=1,
\end{array}                                                        
\label{eq:bangbang}
\end{equation}
along a nonsingular extremal, each control lies at a bound almost
everywhere and can switch only when its score crosses zero. The optimum thus reveals the intuitive control protocol to be a myopic decision that ignores the future: the extremes are ordered in time, and sometimes inverted. Moreover, replacing the backward-propagated terminal gradient by the instantaneous one makes every score nonnegative and recovers the intuitive all-open strategy; a negative score identifies a locally beneficial closing direction for the terminal objective. This is the operational origin of counter-gating: temporarily raising a barrier that instantaneous steepest descent would keep open can improve the terminal outcome.

We stress that the schedules reported here are not constructed from presuming the switching rule: they are obtained by gradient descent over the \emph{full} admissible box $y_e(t)\in[\ymin,1]$, with no bang--bang ansatz imposed. The descent nevertheless drives essentially every gate to a bound, so Eq.~\eqref{eq:bangbang} is an output of the optimization rather than an assumption (End Matter C).

Two features deserve emphasis. (1)~Changing the terminal distance affects the Pontryagin problem through the terminal adjoint.  The dependence on each aperture remains affine, so nonsingular extremals remain bang--bang, although the switching times and best-found protocol may depend on the chosen distance. (2)~Near equilibrium this dependence becomes universal: for any sufficiently smooth \(f\)-divergence with \(f''(1)>0\),
\[
P_\perp\ket{\eta_f(T)}
=
f''(1)\ket{q(T)}
+
O(\|q(T)\|^2),
\]
where \(P_\perp=I-\ket{\sqrt\pi}\!\bra{\sqrt\pi}\). Thus all such divergences share the same leading-order switching structure at large \(T\) \cite{Supplement}.

\emph{Counter-gating in action.---}
For the six-state example, timed gating beats all-open by $130\times$ and the best-found static landscape by $7.05\times$, and the dual schedule retains $2.67\times10^3$ and $1.76$ times as much residual as all-high and the best static landscape (Fig.~\ref{fig:main}). The mechanism is visible in a single number: along the accelerating protocol the same-state contraction-rate ratio
$\mel{n}{L[y]}{n}/\mel{n}{L_{\rm open}}{n}$ falls to $0.31$ and never
exceeds unity. The winning schedule is \emph{nowhere} pointwise faster than all-open, yet it arrives far closer---the literal content of the title. It is also nonmonotone, with $e_2$ opened, closed, and reopened. Under the simple-slow-mode and nonzero-overlap assumptions of End Matter A, counter-gating is forced: any asymptotic speedup over all-open must at least partially close an edge coupled to the all-open slow mode. The target time $T=2.2$ is about $1.6$ slow-mode relaxation times of the all-open dynamics ($\gamma_1^{\rm open}=0.71$).

\begin{figure*}[t]
\includegraphics[width=\textwidth]{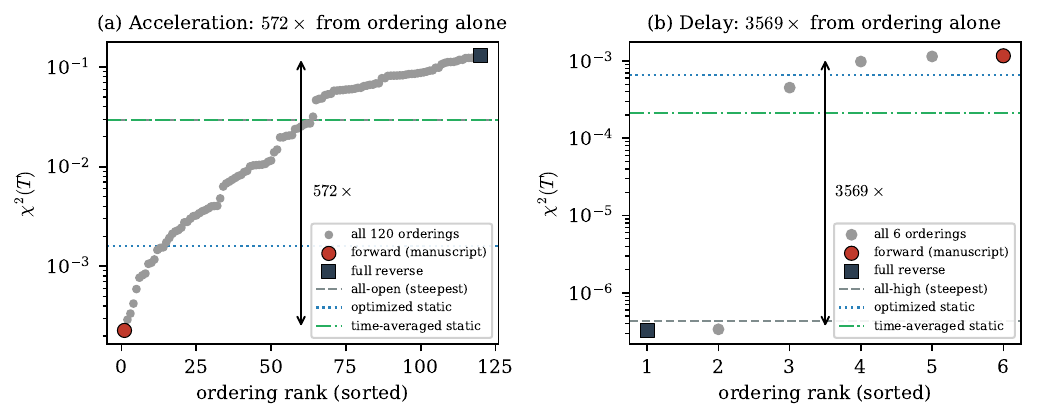}
\caption{\textbf{Same kinetic dose, opposite outcomes: relaxation is
set by the order of gating.}  Each protocol of
Fig.~\ref{fig:main} is reordered by permuting its stages while keeping
every stage's gates and duration fixed, so all orderings share identical
integrated apertures $\int_0^T y_e\,\dd t$ and differ only in time
order. (a) Acceleration: the displayed order gives the smallest
terminal residual and the full reversal the largest among all
$5!=120$ permutations, a ratio of $572$ between the largest and
smallest $\chisq(T)$. (b) Delay: the forward order gives the largest
residual and the reverse order the smallest among all $3!=6$
permutations, a ratio of $3569$. Dashed, dotted, and dash-dotted lines
mark the corresponding pointwise extreme (all-open for acceleration
and all-high for delay), the best-found static landscape, and the
time-averaged static landscape
$\bar y_e=T^{-1}\!\int_0^T y_e\,\dd t$. If the stage generators
commuted, every ordering would have the same endpoint as the static
time-average.}
\label{fig:ordering}
\end{figure*}

\emph{Time ordering as the resource.---}
The cleanest evidence that the time ordering (not the kinetic dosage) is decisive for relaxation speed comes from permuting the stages of either optimal schedule. Holding each stage's gates and duration fixed, so that all orderings share the same integrated gates $\int_0^T y_e\,\dd t$, produces a $572\times$ spread in $\chisq(T)$ for acceleration and a $3569\times$ spread for delay (Fig.~\ref{fig:ordering}), with the forward order best and its reversal worst. The order enters through the commutator: to second order the Magnus expansion \cite{Magnus1954,Blanes2009} of the propagator contains
\begin{equation}
 \Omega_2=\tfrac12\!\int_{t_1>t_2}\![L(t_1),L(t_2)]\,\dd t_1\dd t_2,
 \label{eq:magnus}
\end{equation}
which changes sign under time reversal and vanishes when the generators commute. Independently reoptimized horizon sweeps and a $40$-state uniform-Dirichlet ensemble show the phenomenon is generic rather than fine-tuned: the timed optimum never underperforms the best static landscape, exceeds it by more than $10\%$ for $78\%$ of initial states, with median gain $2.3$ and a long tail (90th percentile $31$). The gain correlates only weakly with slow-mode overlap ($r=0.07$), so overlap alone is not a sufficient statistic for finite-time controllability \cite{Supplement}; the numerical procedures and convergence checks are summarized in End Matter C.

\emph{Thermodynamic accounting and outlook.---}
The forward and reverse problems are both physical: one relaxes as fast as possible, the other preserves as much nonequilibrium free energy as possible at the target time. Both obey the same thermodynamics. The relative entropy satisfies
\begin{equation}
 \dot\KL[\p\Vert\pii]
 =-\sum_{e=\{i,j\}}g_{e}(x_i-x_j)(\ln x_i-\ln x_j)\equiv-\sigma(t), \label{eq:KL}
\end{equation}
and since $\ymin>0$ keeps every generator irreducible, every schedule eventually reaches $\pii$; integrating to $T\to\infty$,
\begin{equation}
\int_0^\infty\sigma(t)\dd t=
\KL[\p_0\Vert\pii],
 \label{eq:invariance}
\end{equation}
the same whether one accelerates, delays, or does nothing. The schedule cannot create free energy or change the eventual relaxation budget; it only changes how much of that budget has been spent by time $T$---front-loaded to accelerate, deferred to delay. Within the reduced Markov description, there is no additional system work, and detailed balance at every instant implies zero housekeeping entropy production. Any energetic cost of implementing the gate modulation belongs to physics outside this reduced description. The reduced description assumes that barrier modulation remains slow relative to intrawell equilibration; violating this separation would excite eliminated modes and introduce additional coarse-graining dissipation outside $\sigma$, a regime the schedules here avoid \cite{Seifert2012}. This contrasts with shortcut-to-adiabaticity constructions that may require effective gain, loss, or non-Markovian generators \cite{Alipour2020}. Here every instantaneous generator is a bona fide reversible Markov generator, and $\chisq$ is a Lyapunov function along \emph{any} schedule.

Under an Arrhenius mapping, the gate ranges used here correspond to barrier excursions of only $1.2$--$3.0\,k_{\rm B}T_{\rm bath}$ \cite{Kramers1940,Hanggi1990}, placing the mechanism within reach of colloidal landscapes, gated reaction networks, and Markov-state descriptions of molecular kinetics \cite{Prinz2011}, where this kinetic freedom is precisely what is already engineered \cite{Kumar2020,Kumar2022,Bechhoefer2021}.

The analysis closes the loop with which we began: the Mpemba-effect literature showed that favorable modal amplitudes, inherited from a special initial state, can beat naive distance. Barrier timing turns that observation into a control principle that no longer relies on initial-state engineering. Mpemba-like acceleration can be synthesized on demand by kinetic scheduling, and, dually, relaxation can be held back---retaining nonequilibrium free energy at the target time---without changing the equilibrium or adding nonconservative driving, simply by the order in which reversible channels are opened or closed. The quantitative limits of such acceleration and retention should be understood alongside stochastic speed-limit constraints \cite{Shiraishi2018,Ito2020}. Finally, the gradient that generates these schedules is itself a response function; because gradients of terminal costs with respect to time-dependent controls can be estimated directly from sampled trajectories via fluctuation--response relations \cite{LyuRayCrutchfield2025}, counter-gating schedules may be discoverable in experiment without a microscopic model of the network.

\paragraph*{Data availability.---}
The data, parameters, and code required to reproduce all numerical
results and figures are openly available in Ref.~\cite{DataArchive}.

\paragraph*{Acknowledgments.---}
The authors appreciate inspiration from Mr. Jinlong Wang, helpful discussions with Prof. Xiaoquan Yu, and a careful reading through of the draft by Mr. Jiming Zheng. This work is supported by the National Science Foundation under Grant No. DMR-2145256 and by the Alfred P. Sloan Foundation under Award No. G-2025-25194.

\bibliography{refs}

\section*{End Matter}
\setcounter{equation}{0}
\renewcommand{\theequation}{A\arabic{equation}}
\renewcommand{\theHequation}{A\arabic{equation}}
\paragraph*{A. No-go theorem and counter-gating certificates.---}
Suppose the $L_e$ are real symmetric, positive semidefinite, and pairwise commuting, they are simultaneously orthogonally diagonalizable, with eigenvalues $\ell_{ea}\ge0$. Writing $c_a(0)=\braket{u_a}{q_0}$, Eq.~\eqref{eq:commuting} gives
\begin{equation}
\chisq(T)=\sum_{a\ge1}|c_a(0)|^2
\exp\!\left[-2T\sum_e\bar y_e\ell_{ea}\right].
\label{eq:commproof}
\end{equation}
Each term is nonincreasing in every $\bar y_e$. Hence all-open globally minimizes, and all-high globally maximizes, the terminal $\chisq$ among all admissible schedules: temporal ordering provides no advantage when the controlled generators commute.

For counter-gating, assume that the all-open slow mode $\ket{u_1}$ is simple and that $c_1(0)=\braket{u_1}{q_0}\ne0$. If every $L[y(t)]$ preserves $\ket{u_1}$, then $\dot c_1=-\lambda_u(t)c_1$, where $\lambda_u(t)=\mel{u_1}{L[y(t)]}{u_1}\le\gamma_1$ because $L[y]\preceq L_{\rm open}$. Thus $|c_1(T)|\ge|c_1(0)|\e^{-\gamma_1T}$, while $\chisq_{\rm open}(T)=|c_1(0)|^2\e^{-2\gamma_1T}[1+o(1)]$; such a protocol therefore cannot asymptotically outperform all-open. Moreover,
$
L[y]\ket{u_1}
=\gamma_1\ket{u_1}
-\sum_e(1-y_e)\bar g_e\braket{v_e}{u_1}\ket{v_e}.
$
Hence mixing the slow mode requires, as a necessary but not sufficient condition, at least one partially closed edge with $y_e<1$ and $\braket{v_e}{u_1}\ne0$. This establishes the necessity of counter-gating for asymptotic acceleration.

A complementary finite-time certificate follows from the first variation about all-open. With $L_0=L_{\rm open}$,
$\ket{q(t)}=\e^{-L_0t}\ket{q_0}$ and
$\ket{\eta(t)}=2\e^{-L_0(2T-t)}\ket{q_0}$, so
\begin{equation}
G_e^{\rm open}(t)=
2\mel{q_0}{\e^{-L_0(2T-t)}L_e\,\e^{-L_0t}}{q_0}.
\label{eq:certificate}
\end{equation}
If $G_e^{\rm open}(t_*)<0$, then slightly closing edge $e$ over a sufficiently short interval around $t_*$ lowers the terminal cost to first order; all-open is therefore not a local optimum. This provides a sufficient acceleration certificate requiring no optimization. If $[L_0,L_e]=0$, then
$
G_e^{\rm open}(t)
=2\mel{q_0}{\e^{-L_0T}L_e\e^{-L_0T}}{q_0}\ge0,
$
the local first-order counterpart of the commuting-control no-go theorem. Thus a negative score certifies noncommutativity and identifies a beneficial counter-gate. For the six-state example,
$\min_{e,t}G_e^{\rm open}=-5.4\times10^{-2}$.

The effect already occurs in the smallest possible dimension. A two-state network has a one-dimensional physical subspace, on which all admissible generators commute, whereas a three-state complete-graph triangle admits a protocol that first holds one barrier high and then opens all, with
$\chisq_{\rm open}(T)/\chisq_{\rm gate}(T)=1.24$.
Parameters and a reproduction script are provided in the Supplemental Material \cite{Supplement}. The six-state network is used to display larger gains, nonmonotone counter-gating, and bidirectional control.

\setcounter{equation}{0}
\renewcommand{\theequation}{B\arabic{equation}}
\renewcommand{\theHequation}{B\arabic{equation}}
\paragraph*{B. Pontryagin variation and moving eigenframe.---}
Adding the dynamical constraint to the endpoint objective and integrating the $\braket{\eta}{\delta\dot q}$ term by parts yields Eq.~\eqref{eq:variation} when Eq.~\eqref{eq:adjoint} holds. At an upper bound, admissible variations satisfy $\delta y_e\le0$; at a lower bound, $\delta y_e\ge0$. Equation~\eqref{eq:bangbang} is the corresponding necessary condition for a nonsingular local minimum and reverses for a local maximum. A singular arc can occur only if $G_e$ vanishes throughout a finite interval; no such interval is resolved in our numerics.

On an interval admitting a differentiable nondegenerate eigenbasis, write $L=U\Gamma U^{\T}$ and $\ket q=U\ket c$. The moving-frame amplitudes obey
\begin{equation}
 \ket{\dot c}=-(\Gamma+A)\ket c,\qquad A=U^{\T}\dot U=-A^{\T}.
 \label{eq:movingframe}
\end{equation}
With $\hat c=\ket c/\|c\|$ and $\Var_{\hat c}(\Gamma)=\mel{\hat c}{\Gamma^2}{\hat c}-\mel{\hat c}{\Gamma}{\hat c}^2$, differentiating $\gamma_{\rm eff}=\mel{\hat c}{\Gamma}{\hat c}$ and using $2\,\mathrm{Re}\mel{\hat c}{\Gamma A}{\hat c}=\mel{\hat c}{[\Gamma,A]}{\hat c}$ yields Eq.~\eqref{eq:threeterm}. At a discontinuous switch, the overlap $U_+^{\T}U_-$ is the finite counterpart of $A$.

\setcounter{equation}{0}
\renewcommand{\theequation}{C\arabic{equation}}
\renewcommand{\theHequation}{C\arabic{equation}}
\paragraph*{C. Numerical methods and verification.---}
All numerics are implemented in Python using NumPy \cite{NumPy2020} and SciPy \cite{SciPy2020}.
Propagation uses symmetric eigendecompositions or direct matrix exponentials, and slice derivatives use the exact divided-difference Fr\'echet formula \cite{Higham2008}; a random finite-difference test gives maximum relative error on the order of $10^{-8}$. Timed schedules are
optimized on grids of $20$ equal-duration slices, the exact gradient of $\chisq(T)$ with respect to every gate on every slice being supplied by the backward adjoint recursion of Eq.~\eqref{eq:adjoint}. The search is a deterministic multistart, box-constrained quasi-Newton descent (L-BFGS-B \cite{Byrd1995}) over the \emph{full} admissible box $y_e\in[\ymin,1]$ on every slice, started from both extreme landscapes, the optimized static control tiled in time, a nearest-horizon continuation, and a random extreme pattern; no
bang--bang ansatz is imposed. The descent nevertheless drives the controls to a bound: of the $160$ control variables, $158$ are saturated at the headline acceleration horizon and all $160$ at the headline delay horizon, and across the target-time sweep the saturated fraction stays above $0.98$ for every $T\le3$ and equals unity for every $T\le1.25$ (acceleration) and every $T\le5$ (delay). Equation~\eqref{eq:bangbang} is therefore an output of the optimization, not an assumption. The five- and three-stage schedules displayed in the figures are distilled from these converged grids by thresholding the few remaining interior values, merging identical consecutive gate patterns, and refining the switch times off the grid. Their switching times are then verified against Pontryagin directly: starting from a $3\%$ random perturbation of the stage durations and holding the stage order fixed, a least-squares shooting solve of the Hamiltonian-jump conditions at the switches \cite{SciPy2020} recovers the displayed durations to relative accuracy $5\times10^{-7}$ and reduces the switching residual norm from $1.3\times10^{-4}$ to $1.8\times10^{-13}$ (acceleration; $2.1\times10^{-6}$
to $3.4\times10^{-16}$ for delay), leaving $\chisq(T)$ unchanged to $3\times10^{-14}$. This is a local refinement at fixed switching structure, not a global search over switching structures. As a further independent check, the $20$-slice reoptimization recovers the displayed gate sequences edge by edge, including the nonmonotone open--high--open excursion of $e_2$, with endpoints within $2\%$ (acceleration) and $0.3\%$ (delay) of the refined ones. Static controls were searched by DIRECT \cite{Jones1993} followed by $80$ multistart L-BFGS-B
refinements, and satisfy the Karush--Kuhn--Tucker conditions with interior gradients below $1.6\times10^{-11}$. The static problem is only eight-dimensional; DIRECT evaluates it $1.2\times10^{4}$ times, and the delay static optimum is a vertex of the control box with strictly positive KKT margins ($2.86\times10^{-5}$). At the longest acceleration horizons the terminal residual falls below $10^{-8}$ and the descent stalls at machine precision before saturating; the horizons reported here are far from that regime. We report optimized and best-found values rather than interval-certified global optima: these checks support numerical credibility but do not certify global optimality.

\end{document}